\font\bbb=msbm10                                    

\overfullrule=0pt

\def\C{\hbox{\bbb C}}

\def\Z{\hbox{\bbb Z}}

\def\FP{{\sl Found.\ Phys.}}

\def\IBMJRD{{\sl IBM J. Res.\ Develop.}}

\def\IJMPC{{\sl Int.\ J. Mod.\ Phys.\ C}}
\def\IJTP{{\sl Int.\ J. Theor.\ Phys.}}

\def\JPA{{\sl J. Phys.\ A:  Math.\ Gen.}}

\def\JSP{{\sl J. Statist.\ Phys.}}

\def\MST{{\sl Math.\ Systems Theory\/}}

\def\PLA{{\sl Phys.\ Lett.\ A\/}}

\def\PLMS{{\sl Proc.\ Lond.\ Math.\ Soc.}}

\def\PRA{{\sl Phys.\ Rev.\ A\/}}

\def\PRD{{\sl Phys.\ Rev.\ D\/}}
\def\PRE{{\sl Phys.\ Rev.\ E\/}}
\def\PRL{{\sl Phys.\ Rev.\ Lett.}}
\def\PRSLA{{\sl Proc.\ Roy.\ Soc.\ Lond.\ A\/}}

\def\Sc{{\sl Science\/}}
\def\SIAMJC{{\sl SIAM J. Comput.}}

\def\dajm{\hbox{D. A. Meyer}}

\def\heather{\hbox{H. Blumer}}
\def\dh{\hbox{\dajm\ and \heather}}

\def\bogtay{\hbox{B. M. Boghosian and W. Taylor, IV}}
\def\bv{\hbox{E. Bernstein and U. Vazirani}}
\def\deutsch{\hbox{D. Deutsch}}
\def\feynman{\hbox{R. P. Feynman}}

\def\grover{\hbox{L. K. Grover}}

\def\shor{\hbox{P. W. Shor}}
\def\simon{\hbox{D. R. Simon}}

\def\hfb{\hfil\break}

\catcode`@=11
\newskip\ttglue

   \font\ninerm=cmr9    \font\eightrm=cmr8   \font\sixrm=cmr6
  \font\ninebf=cmbx9   \font\eightbf=cmbx8  \font\sixbf=cmbx6
  \font\nineit=cmti9   \font\eightit=cmti8  
  \font\ninesl=cmsl9   \font\eightsl=cmsl8  
  \font\ninemi=cmmi9   \font\eightmi=cmmi8  \font\sixmi=cmmi6

\font\bigten=cmr10 scaled\magstep2 

\def\ninepoint{\def\rm{\fam0\ninerm}%
  \textfont0=\ninerm \scriptfont0=\sixrm
  \textfont1=\ninemi \scriptfont1=\sixmi
  \textfont\itfam=\nineit  \def\it{\fam\itfam\nineit}%
  \textfont\slfam=\ninesl  \def\sl{\fam\slfam\ninesl}%
  \textfont\bffam=\ninebf  \scriptfont\bffam=\sixbf
    \def\bf{\fam\bffam\ninebf}%
  \tt \ttglue=.5em plus.25em minus.15em
  \normalbaselineskip=11pt
  \setbox\strutbox=\hbox{\vrule height8pt depth3pt width0pt}%
  \normalbaselines\rm}

\def\eightpoint{\def\rm{\fam0\eightrm}%
  \textfont0=\eightrm \scriptfont0=\sixrm
  \textfont1=\eightmi \scriptfont1=\sixmi
  \textfont\itfam=\eightit  \def\it{\fam\itfam\eightit}%
  \textfont\slfam=\eightsl  \def\sl{\fam\slfam\eightsl}%
  \textfont\bffam=\eightbf  \scriptfont\bffam=\sixbf
    \def\bf{\fam\bffam\eightbf}%
  \tt \ttglue=.5em plus.25em minus.15em
  \normalbaselineskip=9pt
  \setbox\strutbox=\hbox{\vrule height7pt depth2pt width0pt}%
  \normalbaselines\rm}

\def\sfootnote#1{\edef\@sf{\spacefactor\the\spacefactor}#1\@sf
      \insert\footins\bgroup\eightpoint
      \interlinepenalty100 \let\par=\endgraf
        \leftskip=0pt \rightskip=0pt
        \splittopskip=10pt plus 1pt minus 1pt \floatingpenalty=20000
        \parskip=0pt\smallskip\item{#1}\bgroup\strut\aftergroup\@foot\let\next}
\skip\footins=12pt plus 2pt minus 2pt
\dimen\footins=30pc

\def\ie{{\it i.e.}}
\def\eg{{\it e.g.}}

\def\Example{E{\eightpoint XAMPLE}}

\def\xor{{\eightpoint XOR}}

\def\and{{\eightpoint AND}}

\def\Htwon{(\C^2)^{\otimes n}}

\def\looongrightarrow{\relbar\joinrel\relbar\joinrel\relbar\joinrel\rightarrow}
\def\looongmapsto{\mapstochar\looongrightarrow}

\def\Feynman{1}
\def\Zalka{2}
\def\Wiesner{3}
\def\Lloyd{4}
\def\BTsim{5}
\def\AbramsLloyd{6}
\def\FKW{7}
\def\OGKL{8}
\def\Shor{9}
\def\LidarBiham{10}
\def\Yepez{11}
\def\pqa{12}
\def\NielsenChuang{13}
\def\qcaqlg{14}
\def\Bennett{15}
\def\Toffoli{16}
\def\ADH{17}
\def\DeutschJozsa{18}
\def\Simon{19}
\def\BernsteinVazirani{20}
\def\Grover{21}
\def\vanDam{22}
\def\qsa{23}
\def\BTmulti{24}
\def\lgbu{25}
\def\qlgaI{26}
\def\gauge{27}
\def\qlgaII{28}
\def\Taylor{29}
\def\pglga{30}
\def\nogo{31}
\def\Feynmanlect{32}
\def\FarhiGutmann{33}
\def\CFG{34}
\def\AAKV{35}
\def\ABNVW{36}
\def\Coopersmith{37}
\def\DiVincenzo{38}

\magnification=1200

\input epsf.tex

\dimen0=\hsize \divide\dimen0 by 13 \dimendef\chasm=0
\dimen1=\chasm \multiply\dimen1 by  6 \dimendef\halfwidth=1
\dimen2=\chasm \multiply\dimen2 by  7 \dimendef\secondstart=2
\dimen3=\chasm \divide\dimen3 by 2 \dimendef\quarter=3
\dimen4=\quarter \multiply\dimen4 by 9 \dimendef\twopointtwofivein=4
\dimen5=\chasm \multiply\dimen5 by 3 \dimendef\onepointfivein=5
\dimen6=\chasm \multiply\dimen6 by 7 \dimendef\threepointfivein=6
\dimen7=\hsize \advance\dimen7 by -\chasm \dimendef\usewidth=7
\dimen8=\chasm \multiply\dimen8 by 4 \dimendef\thirdwidth=8
\dimen9=\usewidth \divide\dimen9 by 2 \dimendef\halfwidth=9

\line{\hfill                                           12 August 2001}
\line{\hfill                                   {\tt quant-ph/0111069}}
\vfill
\centerline{\bf\bigten QUANTUM COMPUTING CLASSICAL PHYSICS}
\bigskip\bigskip
\centerline{\bf David A. Meyer}
\bigskip
\centerline{\sl Project in Geometry and Physics}
\centerline{\sl Department of Mathematics}
\centerline{\sl University of California/San Diego}
\centerline{\sl La Jolla, CA 92093-0112 USA}
\centerline{\tt dmeyer@chonji.ucsd.edu}
\vfill
\centerline{ABSTRACT}
\bigskip
\noindent In the past decade quantum algorithms have been found which 
outperform the best classical solutions known for certain classical 
problems as well as the best classical methods known for simulation of 
certain quantum systems.  This suggests that they may also speed up 
the simulation of some {\sl classical\/} systems.  I describe one 
class of discrete quantum algorithms which do so---quantum lattice gas 
automata---and show how to implement them efficiently on standard 
quantum computers.

\bigskip\bigskip
\noindent 2001 Physics and Astronomy Classification Scheme:
                   03.67.Lx. 
                   05.10.-a. 

\noindent 2000 American Mathematical Society Subject Classification:
                   81P68,    
                   65Z05.    

\smallskip
\global\setbox1=\hbox{Key Words:\enspace}
\parindent=\wd1
\item{Key Words:}  quantum lattice gas automata, quantum simulation,
                   quantum Fourier transform.

\vfill
\hrule width2.0truein
\medskip
\noindent Expanded version of an invited talk presented at the NATO
Advanced Workshop {\sl Discrete Simulation of Fluid Dynamics:  New 
Trends, New Perspectives}, Carg\`ese, France, 2--6 July 2001.
\eject

\headline{\ninepoint\it Quantum computing classical physics
                                                \hfill David A. Meyer}

\parskip=10pt
\parindent=20pt

\noindent{\bf 1.  Introduction}

\noindent Quantum computing originated with Feynman's observation that
quantum systems are hard to simulate on classical computers, but that
they might be easier to simulate if one had a computer which could be
operated quantum mechanically [\Feynman].  Developments during the 
subsequent two decades have not only supported this observation
[\Zalka--\OGKL], but have also 
demonstrated that quantum computers would---if they existed---solve 
certain {\sl classical\/} problems like factoring more efficiently 
than is possible on classical computers running the best classical 
algorithms known [\Shor].  This raises a natural question 
[\LidarBiham--\pqa]:  Might quantum computers 
help with the simulation of classical systems?  Or more specifically, 
given the focus of this workshop, could quantum computers efficiently 
simulate fluid dynamics?  Without answering the specific question, in 
this paper I try to explain why the answer to the general question may 
be `yes', and along the way explain some of the quantum algorithmic 
tricks which seem likely to be useful in future investigations of 
these questions.

I begin by (very) rapidly introducing the ideas of quantum computation
in \S2.  For a complete presentation, see [\NielsenChuang].  In \S3 I 
describe one approach to simulating quantum systems---quantum lattice
gas automata (QLGA) [\qcaqlg]---and then in \S4 explain a connection 
with simulation of classical systems.  The crucial issue is the 
relative complexity of quantum and classical algorithms; so \S5 
contains a detailed analysis for one specific problem, including some 
new results on implementing QLGA algorithms on `standard' quantum 
computers.  I conclude with a brief discussion in \S6.

\medskip
\noindent{\bf 2.  Quantum computers}

\noindent The possible states of a classical computer are (very long)
bit strings $b_1\ldots b_n \in \{0,1\}^n$.  A particular computation 
proceeds {\it via\/} a sequence of maps to new bit strings:  
$b'_1\ldots b'_n$, $\ldots$.  A fundamental result, which contributed 
directly to the conceptual development of quantum computation, is that 
any classical computation can be made to be {\sl reversible\/} 
[\Bennett], \ie, these maps can be chosen to be {\sl permutations\/} 
on the space of states.  Quantum computation can then be understood as 
a generalization of classical computation:  The possible states of a 
quantum computer are superpositions of bit strings 
$\sum a_{b_1\ldots b_n} |b_1\ldots b_n\rangle \in \Htwon$ (each $\C^2$ 
tensor factor is called a quantum bit or {\sl qubit\/}), where 
$\sum |a_{b_1\ldots b_n}|^2 = 1$ so that the norm-squared of each 
amplitude $a_{b_1\ldots b_n} \in \C$ is the probability that the state 
$|b_1\ldots b_n\rangle$ is observed if the quantum system is measured 
in this `computational basis'.  A particular quantum computation 
proceeds {\it via\/} a sequence of {\sl unitary\/} maps to new states 
$\sum a'_{b_1\ldots b_n} |b_1\ldots b_n\rangle$, $\ldots$.  This much 
is a generalization of classical reversible computation since 
permutations are unitary maps, and each classical state is an allowed 
quantum state.  The difference is that the final state is not directly 
available; it can only be sampled according to the probabilities given 
by the norm-squared of the amplitudes.

To evaluate the computational complexity of an algorithm, either 
classical or quantum, we must specify a set of elementary operations, 
the number of which used during the computation quantifies the
complexity.  If we allow arbitrary permutations classically, or 
arbitrary unitary transformations quantum mechanically, any state can
be reached from\break

\null\vskip-2\baselineskip
\moveright\secondstart\vtop to 0pt{\hsize=\halfwidth
\vskip -2\baselineskip
$$
\epsfxsize=\halfwidth\epsfbox{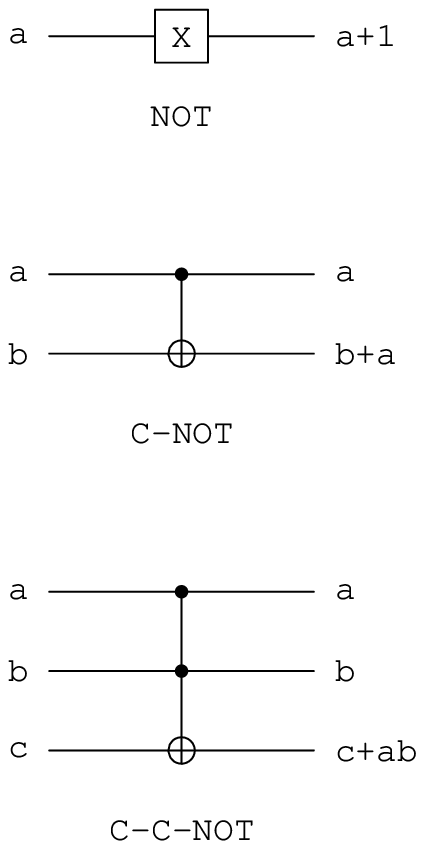}
$$
\vskip -0.75\baselineskip
\eightpoint{%
\noindent{\bf Figure~1}.  A universal set of classical reversible
gates.  Inputs are on the left; outputs are on the right.  $+$ denotes
addition mod 2.
}}
\vskip -\baselineskip
\parshape=19
0pt \halfwidth
0pt \halfwidth
0pt \halfwidth
0pt \halfwidth
0pt \halfwidth
0pt \halfwidth
0pt \halfwidth
0pt \halfwidth
0pt \halfwidth
0pt \halfwidth
0pt \halfwidth
0pt \halfwidth
0pt \halfwidth
0pt \halfwidth
0pt \halfwidth
0pt \halfwidth
0pt \halfwidth
0pt \halfwidth
0pt \hsize
\noindent any other state in a single step---these are clearly
not reasonable models of computation.  Instead, consider the `gate
operations' shown in Fig.~1:  {\eightpoint NOT}, {\eightpoint C-NOT} 
and {\eightpoint C-C-NOT} (`{\eightpoint C}' abbreviates 
`{\eightpoint CONTROLLED}').  Each of these is a permutation on the 
set of bit strings, and Toffoli [\Toffoli] has shown that this set of 
three gates is {\sl universal\/} for classical computation, in the 
sense that any (reversible) boolean operation can be decomposed as a 
sequence of these gates---which by Bennett's result [\Bennett] 
suffices for universality.  Each of these gates can be extended by 
linearity to a unitary map on $\Htwon$, acting non-trivially only on a 
subset of 1, 2, or 3 qubits, and thus can also be used as a 
{\sl quantum\/} gate.  (Traditionally the quantum {\eightpoint NOT} 
gate is denoted by $\sigma_x$ or $X$.)  Two other quantum gates which 
are particularly useful act on single qubits:
$$
H = {1\over\sqrt{2}}\pmatrix{ 1 &  1 \cr
                              1 & -1 \cr
                            }
\quad{\rm and}\quad
R_{\omega} = \pmatrix{ 1 &    0   \cr
                       0 & \omega \cr
                     },
$$
the `Hadamard transform' and phase $\omega$ rotation, respectively.
These matrices have been expressed in the computational basis; thus
$$
\eqalign{
H|0\rangle &\mapsto {1\over\sqrt{2}}(|0\rangle + |1\rangle)        \cr
H|1\rangle &\mapsto {1\over\sqrt{2}}(|0\rangle - |1\rangle).       \cr
}
$$
{\eightpoint C-NOT} and $R_{\omega}$ for $\omega = e^{i\theta}$ with 
$\theta = \cos^{-1}{3\over5}$ form a universal set of gates for 
quantum computation [\ADH].

As we noted earlier, any array of (reversible) classical gates can be
simulated by some array of quantum gates.  The remarkable fact is that
in some cases fewer gates are required quantum mechanically.  The 
following example is the smallest version of the Deutsch-Jozsa 
[\DeutschJozsa] and Simon [\Simon] problems:

\Example.  Given a function $f : \{0,1\} \to \{0,1\}$, we would like 
to evaluate $f(0)$ \xor\ $f(1)$.  The function is accessed by calls 
which have the effect of taking classical states $(x,b)$ to 
$\bigl(x,b\oplus f(x)\bigr)$ where $\oplus$ denotes addition mod~2.
This is a reversible operation and thus also defines a unitary 
transformation on a pair of qubits:  $f$-{\eightpoint C-NOT}.  
Classically, this gate must be applied at least twice (once with 
$x = 0$ and once with $x = 1$) in any algorithm which outputs $f(0)$ 
\xor\ $f(1)$ correctly with probability greater than ${1\over2}$ 
(assuming a uniform distribution on the possible functions).  Quantum 
mechanically we can exploit interference to do better, applying the
operation only once.  Suppose the system is initialized in the state
$|0\rangle\otimes|0\rangle$.  Then apply the following sequence of 
unitary operations:
$$
\def\mapstoright#1{\smash{\mathop{\looongmapsto}\limits^{#1}}}
\eqalign{
|0\rangle\otimes|0\rangle\; 
\mapstoright{H\otimes HX}\;
 & \sum_{x=0}^1 {1\over2}|x\rangle\otimes(|0\rangle - |1\rangle)  \cr
\mapstoright{f{\scriptscriptstyle\rm CNOT}}\;
 & \sum_{x=0}^1 {1\over2} (-1)^{f(x)}
                |x\rangle\otimes(|0\rangle - |1\rangle)           \cr
\mapstoright{H\otimes I_2}\;
 & \sum_{x,y=0}^1 {1\over2\sqrt{2}} (-1)^{f(x)} (-1)^{xy}
                  |y\rangle\otimes(|0\rangle - |1\rangle)         \cr
=\;
 & {1\over\sqrt{2}} |f(0) \hbox{\ \xor\ } f(1)\rangle\otimes 
                           (|0\rangle - |1\rangle).               \cr
}
$$
The first step Fourier transforms the `query' qubit into an equal 
superposition of $|0\rangle$ and $|1\rangle$, and initializes the
`response' qubit into a state which will create the phase 
$(-1)^{f(x)}$ in the second step.  The third step Fourier transforms
the query qubit again, creating the interference which ensures that
subsequently measuring the query qubit outputs $f(0)$ \xor\ $f(1)$ 
correctly with probability 1.  The gate array implementing this 
quantum algorithm is shown in Fig.~2; it includes only one 
$f$-{\eightpoint C-NOT} gate.

\midinsert
\null\vskip-2\baselineskip
$$
\epsfxsize=\threepointfivein\epsfbox{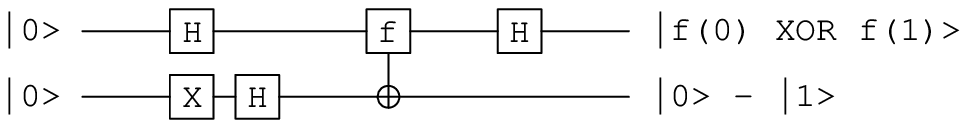}
$$
\vskip-\baselineskip
\eightpoint{%
{\narrower\noindent{\bf Figure~2}.  The gate array implementing the
quantum algorithm which solves the {\sixrm XOR} problem with a single 
call to the $f$-{\sixrm C-NOT} gate.  The states are not 
normalized.\par
}}
\endinsert

This example displays perhaps the simplest quantum improvement over 
classical computation.  Other superior quantum solutions to classical 
problems include Deutsch and Jozsa's balanced function algorithm 
[\DeutschJozsa], Bernstein and Vazirani's parity search algorithm 
[\BernsteinVazirani], Simon's period finding algorithm [\Simon], 
Shor's factoring algorithm [\Shor], Grover's unstructured search 
algorithm [\Grover], van Dam's algorithms for weighing matrices and
quadratic residues [\vanDam], and Hunziker and Meyer's highly 
structured quantum search algorithms [\qsa].

\medskip
\noindent{\bf 3.  Quantum lattice gas automata}

\noindent Although the example in \S2 demonstrates the superior
computational power of quantum systems for certain problems, it seems
to have little to do with confirming Feynman's original idea that 
quantum systems could efficiently simulate other quantum systems 
[\Feynman].  For the community attending this workshop, a natural 
place to look for such confirmation is lattice models:  The possible 
configurations for each particle on a one dimensional lattice $L$ are 
labelled by pairs $(x,\alpha) \in L \times \{\pm1\}$, where $x$ is the 
position and $\alpha$ the velocity.  A classical lattice gas evolution 
rule consists of an advection stage
$(x,\alpha) \mapsto (x+\alpha,\alpha)$, followed by a scattering 
stage.  Each particle in a quantum lattice gas automaton (QLGA) 
[\qcaqlg] exists in states which are superpositions of the classical 
states:  $|\psi\rangle = \sum\psi_{x,\alpha} |x,\alpha\rangle$, where 
$1 = \langle\psi|\psi\rangle 
   = \sum \bar\psi_{x,\alpha} \psi_{x,\alpha}$.  The evolution rule 
must be unitary; the most general with the same form as the classical 
rule is:
$$
\def\mapstoright#1{\smash{\mathop{\looongmapsto}\limits^{#1}}}
\eqalign{
\sum\psi_{x,\alpha} |x,\alpha\rangle\;
\mapstoright{\rm advect}\;
 & \sum\psi_{x,\alpha} |x+\alpha,\alpha\rangle                      \cr
\mapstoright{\rm scatter}\;
 & \sum\psi_{x,\alpha} S_{\alpha\alpha'} |x+\alpha,\alpha'\rangle,  \cr
}
$$
where the scattering matrix is
$$
S = \pmatrix{ \cos s & i\sin s \cr
             i\sin s &  \cos s \cr
            }.
$$

\moveright\secondstart\vtop to 0pt{\hsize=\halfwidth
\vskip -2.2\baselineskip
$$
\epsfxsize=\halfwidth\epsfbox{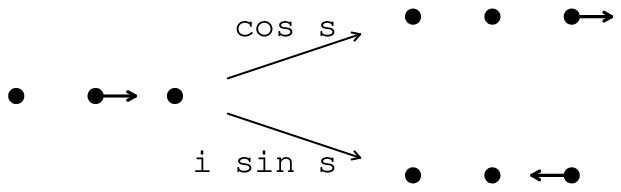}
$$
\vskip -0.75\baselineskip
\eightpoint{%
\noindent{\bf Figure~3}.  The general evolution rule for a single 
particle in the one dimensional QLGA.
}}
\vskip -\baselineskip
\parshape=8
0pt \halfwidth
0pt \halfwidth
0pt \halfwidth
0pt \halfwidth
0pt \halfwidth
0pt \halfwidth
0pt \halfwidth
0pt \hsize
\noindent Let $U$ denote the complete single timestep evolution 
operator for a single particle, the composition of advection and 
scattering.  Fig.~3 illustrates this quantum evolution:  at $s = 0$ 
it specializes to the classical deterministic lattice gas rule.  The 
$\Delta x = \Delta t \to 0$ limit of this discrete time evolution is 
the Dirac equation [\qcaqlg]; the $\Delta x^2 = \Delta t \to 0$ limit 
is the Schr\"odinger equation [\BTmulti].

QLGA models can be generalized to higher dimensions [\BTmulti], and to
include more particles [\qcaqlg,\lgbu,\BTmulti], potentials 
[\qlgaI,\gauge] and various boundary conditions [\qlgaII].  These are
quantum models which we might try to simulate classically or quantum
mechanically.  Figures~4--6 show the results of classical simulations
of plane waves, wave packets, and scattering off potential steps, 
respectively.  These support the claim in the previous paragraph that
QLGA are discrete models for quantum particles.  In the next section 
we will see how they are also relevant to the question of simulating
classical physical systems quantum mechanically.

\midinsert
\null\vskip-\baselineskip
\vskip-\baselineskip
$$
\epsfxsize=\halfwidth\epsfbox{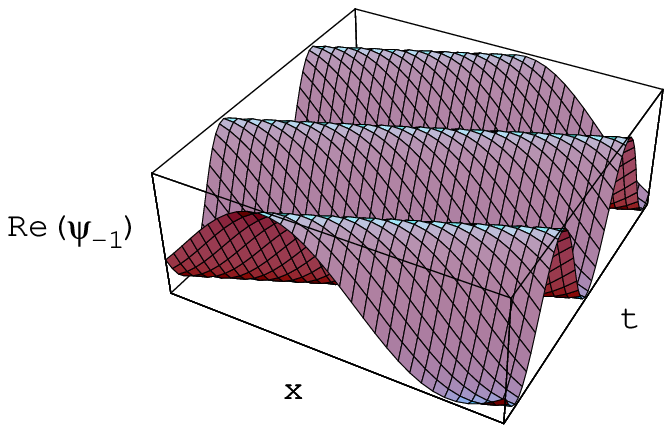}\hskip\chasm%
\epsfxsize=\halfwidth\epsfbox{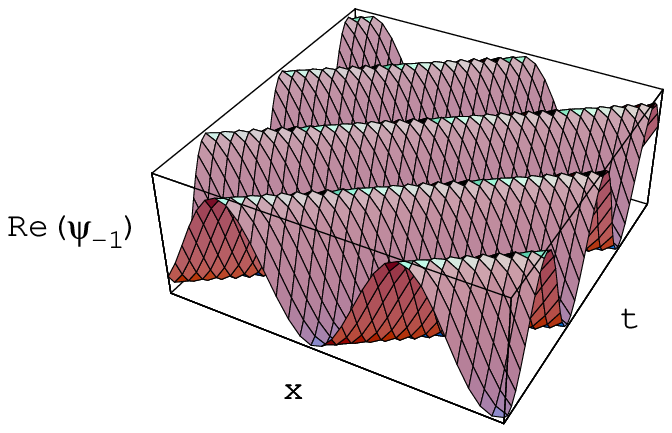}
$$
\eightpoint{%
\centerline{{\bf Figure~4}.  Plane waves in the general one
dimensional QLGA [\qlgaI].}}
\endinsert

\pageinsert
\null\vskip-3\baselineskip
$$
\epsfxsize=\usewidth\epsfbox{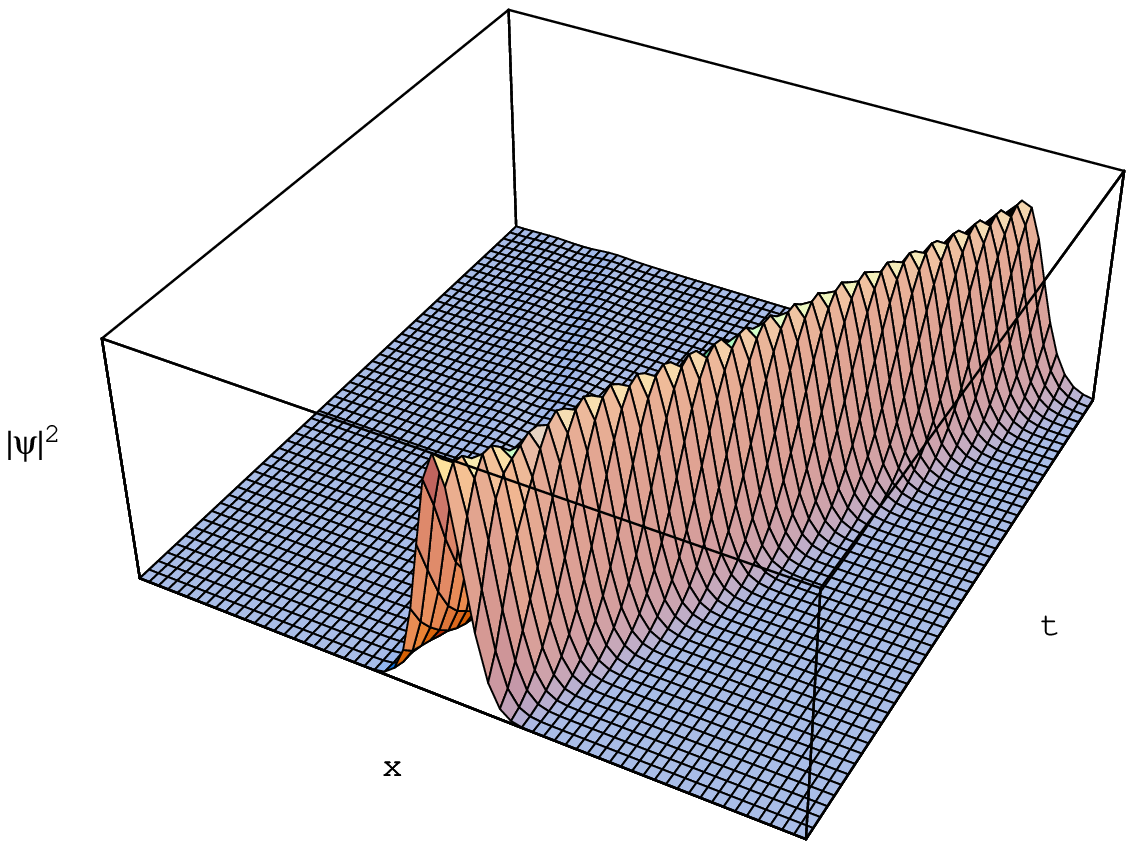}
$$
\vskip-\baselineskip
\eightpoint{%
{\narrower\noindent{\bf Figure 5}.  Evolution of a wave packet in the
general one dimensional QLGA [\qlgaI].\par} 
}

\vfill
\null\vskip-8\baselineskip
$$
\epsfxsize=\usewidth\epsfbox{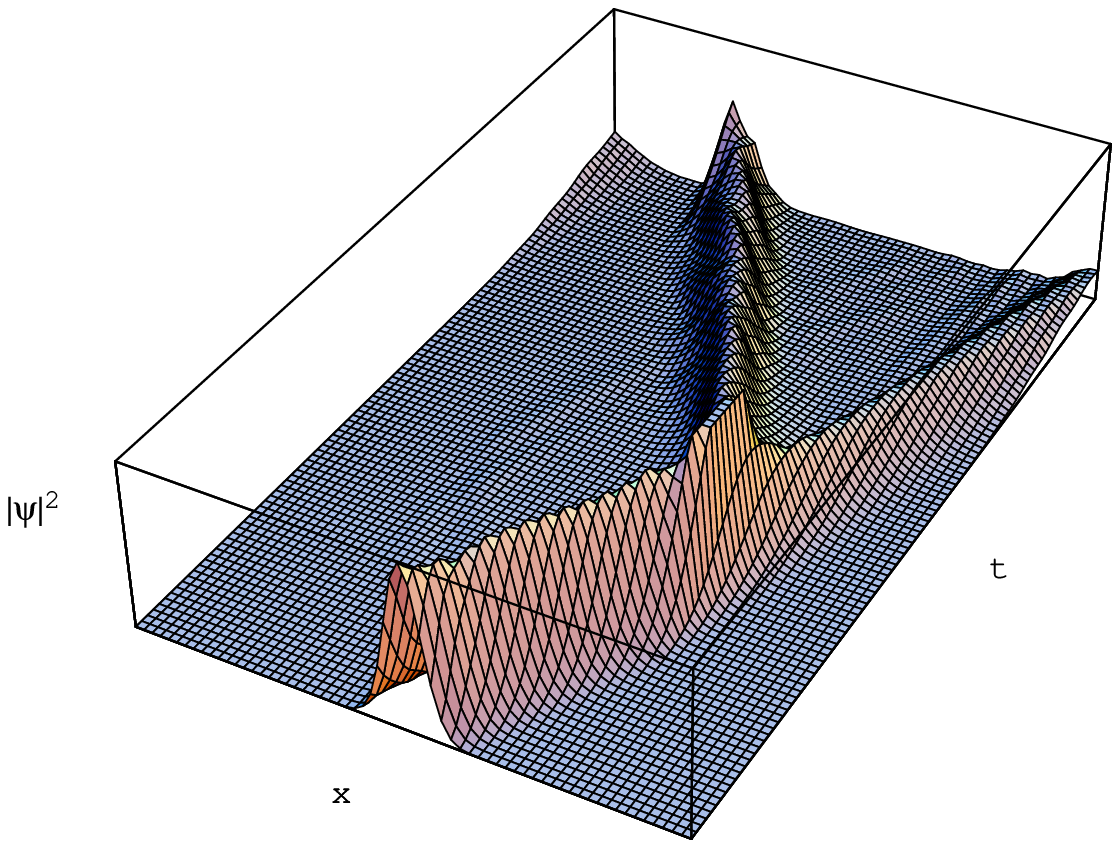}
$$
\vskip-8\baselineskip
\eightpoint{%
{\narrower\noindent{\bf Figure 6}.  Scattering of a wave packet from a
potential step in the general one dimensional QLGA [\qlgaI].\par} 
}
\endinsert

\vfill\eject
\medskip
\noindent{\bf 4.  Diffusion}
\nobreak

\nobreak
\noindent The evolution rule for a single particle QLGA bears some 
resemblence to a random walk.  More precisely, it is the unitary 
version of a correlated random walk [\Taylor,\pglga]---with 
$\theta = \pi/4$ the analogue of uncorrelated.  In Fig.~5, for 
example, we can see evolution like that of a biased random walk, with
a spreading Gaussian distribution.  And, in fact, this is as close as
possible---there is no quantum random walk in the sense of local,
solely $x$ dependent unitary evolution; the $\alpha$ dependence must
be included [\nogo].  (This is not true for quantum processes 
{\sl continuous\/} in time [\Feynmanlect], which can also be used for
computational purposes like the ones we are considering here 
[\FarhiGutmann,\CFG].)

Consequently, there are differences.  Diffusion approaches an 
equilibrium state, independently of the initial condition (on 
connected spaces).  Unitary evolution of a single particle QLGA 
cannot:  the distance $\delta$ between successive states, defined by 
$\cos\delta = \langle U\psi|\psi\rangle$ is constant, so the evolution
of any state which is not an eigenstate of $U$ with eigenvalue 1 does
not converge.  Each state implies a probability distribution on the
lattice, given by
$$
P_t(x) = {\rm prob}(x;t) = |\psi_{x,-1}(t)|^2 + |\psi_{x,+1}(t)|^2,
$$
so we can also ask if this converges.  In fact, this probability 
distribution is constant for each eigenstate $|\phi_i\rangle$ of $U$
[\qlgaI].  But since there exists $T \in \Z_{>0}$ such that 
$\lambda_i^T \sim 1$ for all eigenvalues $\lambda_i$ of $U$, and hence
$U^T \psi \sim \psi$, for any initial state such that $P_1 \not= P_0$,
the probability distribution cannot converge either [\AAKV].

Aharonov, Ambainis, Kempe and Vazirani have shown, however, that the
time average of the probability distribution does converge [\AAKV]:
Expand the initial state $|\psi\rangle = \sum_i a_i|\phi_i\rangle$ in 
terms of the eigenvectors of $U$.  Then 
$U^t|\psi\rangle = \sum_i a_i^{\vphantom t}\lambda_i^t|\phi_i\rangle$, 
so
$$
\eqalign{
{\rm prob}(x,\alpha;t)
 &= \Bigl|\sum_i a_i^{\vphantom t} \lambda_i^t 
          \langle x,\alpha|\phi_i\rangle
    \Bigr|^2                                                       \cr
 &= \sum_{i,j} a_i \bar a_j (\lambda_i \bar\lambda_j)^t
               \langle x,\alpha|\phi_i\rangle 
               \langle \phi_j|x,\alpha\rangle.                     \cr
}
$$
Then the time average of the probability is 
$$
{1\over T}\sum_{t=0}^{T-1} {\rm prob}(x,\alpha;t)
 =
\sum_{i,j} a_i \bar a_j \langle x,\alpha|\phi_i\rangle 
                        \langle \phi_j|x,\alpha\rangle
           {1\over T} \sum_{t=0}^{T-1} (\lambda_i \bar\lambda_j)^t.
$$
For $\lambda_i \not= \lambda_j$, the interior sum goes to 0  as 
$T \to \infty$.  This leaves only the terms in the sum for which
$\lambda_i = \lambda_j$, which are independent of $T$.  Thus the time
average converges.  In particular, for the one dimensional single 
particle QLGA, it converges to the uniform distribution which is the 
equilibrium distribution for diffusion on one dimensional lattices.  
That is, by measuring the position at random times we can simulate 
sampling from the equilibrium distribution of classical diffusion.  
Although this is not true for all graphs, \eg, the Cayley graph of the
{\sl nonabelian\/} group $S_3$ [\AAKV], we have analyzed one example 
of discrete quantum simulation of a classical physical process.

\medskip
\noindent{\bf 5.  Computational complexity}
\nobreak

\nobreak
\noindent To be truely useful, the quantum computation should be more
efficient than the corresponding classical computation.  Classically,
$O(N^2)$ steps of a random walk are required to approximate the
equilibrium distribution on a lattice of size $N$.  Aharonov {\it et
al.}\ have shown that only $O(N\log N)$ steps of the single particle 
QLGA are required for equally accurate sampling [\AAKV], and more 
detailed calculations by Ambainis, Bach, Nayak, Vishwanath and Watrous 
[\ABNVW] show that $O(N)$ steps suffice.  The proofs of these results 
depend on careful estimates about, for example, the distribution of 
eigenvalues of $U$ and are somewhat involved.  Simple simulations, 
however, provide a heuristic explanation for this quantum improvement.  
Fig.~7 shows the evolution of a QLGA particle initialized at 
$|\psi\rangle = (|0,-1\rangle + |0,+1\rangle)/\sqrt{2}$.  Notice that
the peaks of the probability distribution---indicated by the darkest
squares in the plot---spread approximately linearly in time.  This is
the origin of $O(N)$ number of steps required for the QLGA to sample 
the equilibrium distribution.

\pageinsert
$$
\epsfxsize=\thirdwidth\epsfbox{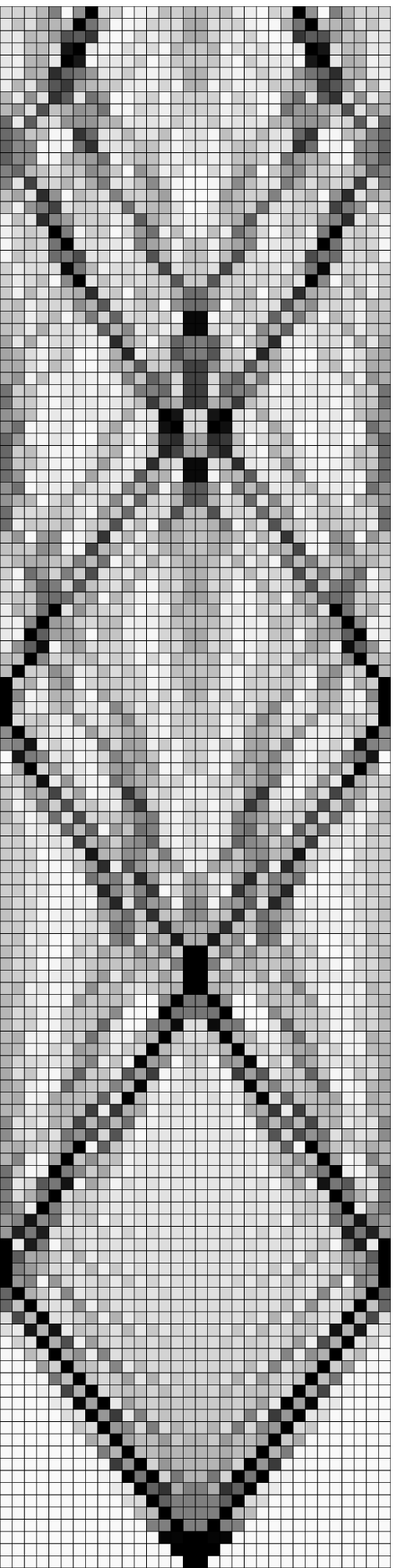}
$$
\vskip-\baselineskip
\eightpoint{%
{\narrower\noindent{\bf Figure 7}.  Evolution of a single particle
QLGA initialized at $x=0$ in an equal superposition of left and right
moving states [\qcaqlg].  Each square corresponds to one of the 
classical states; thus each lattice site is represented by {\sl two\/}
squares, one for each allowed velocity.  The boundary conditions are
periodic and time runs upward.\par} 
}
\endinsert

Thus the QLGA provides a quadratic improvement in the number of steps
required compared to a classical random walk.  To verify a 
computational improvement, however, the QLGA must not require much 
more computation per step.  Each step of the random walk requires a 
coin-flip, \ie, generation of a random number.  Inasmuch as this is
possible classically, it requires {\sl constant\/} time, independent 
of $N$.  Each step also requires the addition of the $\pm1$ result of
the coin flip to the current position.  Since the latter is a $\log N$
bit integer, this requires $O(\log N)$ elementary operations.  Thus
the total number of computational steps to simulate one run of the
random walk is $O(N^2\log N)$.  We can compare this with the 
computation required to compute the whole probability distribution by
evaluating the Markov process.  Although this could be computed by 
matrix multiplication at each step, the locality of the process means
that we need only compute
$$ 
{\rm prob}(x;t+1) = {1\over2} {\rm prob}(x-1;t) 
                  + {1\over2} {\rm prob}(x+1;t)
$$ 
for each lattice point.  Since this has a constant computational cost 
per lattice point, evaluating the whole probability distribution for 
$O(N^2)$ steps takes $O(N^3)$ elementary operations.

For the QLGA we cannot run single trajectories since that would miss 
any interference between trajectories.  Thus classical simulation
of the QLGA must be like the Markov process calculation of the whole
probability distribution of the random walk.  Again, by locality, each
step requires constant computation per lattice point.  Thus evolution 
of the whole state for $O(N)$ steps takes $O(N^2)$ elementary 
operations.  Taking the time average to approximate the diffusive 
equilibrium distribution requires another $O(N)$ factor, hence the 
same $O(N^3)$ elementary operations we found in the previous 
paragraph.  Unsurprisingly, therefore, we have not discovered a faster 
{\sl classical\/} algorithm for simulating the equilibrium 
distribution of diffusion.

\moveright\secondstart\vtop to 0pt{\hsize=\halfwidth
\vskip -1\baselineskip
$$
\epsfxsize=\halfwidth\epsfbox{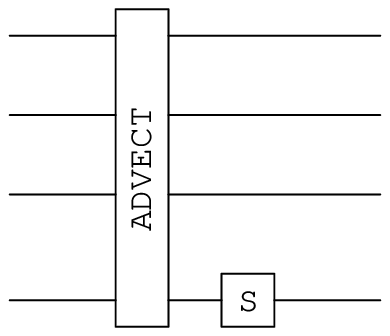}
$$
\vskip -0.75\baselineskip
\eightpoint{%
\noindent{\bf Figure~8}.  The high level circuit for $U$.  There are
$\log N$ qubits labelling the position, and a last qubit encoding the
velocity.
}}
\vskip -\baselineskip
\parshape=11
0pt \halfwidth
0pt \halfwidth
0pt \halfwidth
0pt \halfwidth
0pt \halfwidth
0pt \halfwidth
0pt \halfwidth
0pt \halfwidth
0pt \halfwidth
0pt \halfwidth
0pt \hsize
Thus, to realize a {\sl quantum mechanical\/} improvement we must be 
able to compute a single step $U$ of the QLGA evolution with fewer 
than $O(N)$ elementary operations on a quantum computer.\break
Schematically, $U$ is implemented as shown in Fig.~8, acting on 
$\log N$ qubits which encode the position on the lattice and a single
qubit which encodes the velocity. The scattering operation $S$ acts on
the last qubit and thus has constant cost per step.  Advection is a 
shift operation, \ie, multiplication by a matrix which is 
diagonalized by the discrete Fourier transform $F_N$.  The left shift
$|x\rangle \mapsto |x-1\rangle$ is
$$
\pmatrix{   0   &   1   &       &       &       \cr
                &   0   &   1   &       &       \cr
                &       & \ddots& \ddots&       \cr
                &       &       &   0   &   1   \cr
            1   &       &       &       &   0   \cr
        }
 = 
F_N 
\pmatrix{ 1 &          &          &          &              \cr
            &  \omega  &          &          &              \cr
            &          & \omega^2 &          &              \cr
            &          &          &  \ddots  &              \cr
            &          &          &          & \omega^{N-1} \cr
        }
F_N^{\dagger}.                                                \eqno(1)
$$
Here $\omega = e^{2\pi i/N}$ and the right shift diagonalizes to the 
same matrix using $F_N^{\dagger}$ rather than $F_N$.  Assuming 
$N = 2^n$, the diagonal matrix can be implemented by $n = \log N$ 
single qubit phase rotations:
$R_{\omega} \otimes R_{\omega^2} \otimes R_{\omega^4} \otimes \cdots
            \otimes R_{\omega^{N/2}}$.  

\moveright\secondstart\vtop to 0pt{\hsize=\halfwidth
\vskip -1\baselineskip
$$
\epsfxsize=\halfwidth\epsfbox{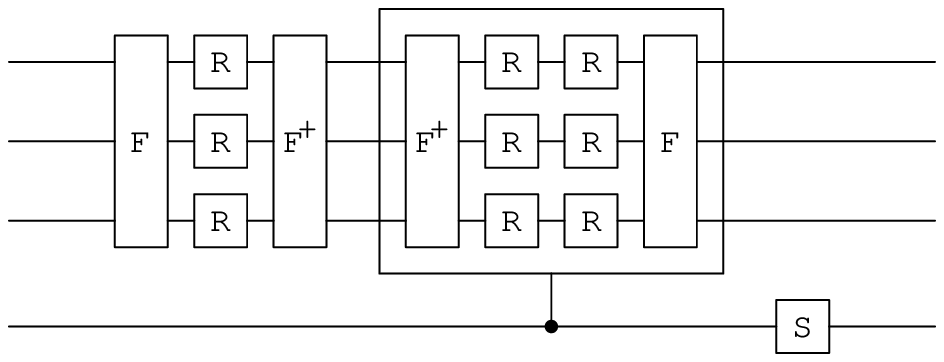}
$$
\vskip -0.75\baselineskip
\eightpoint{%
\noindent{\bf Figure~9}.  A low level implementation of $U$.  The 
phase rotations are as in the text, namely 
$R_{\omega} \otimes R_{\omega^2} \otimes R_{\omega^4}$ in each of the
three timesteps at which they are applied.
}}
\vskip -\baselineskip
\parshape=13
0pt \halfwidth
0pt \halfwidth
0pt \halfwidth
0pt \halfwidth
0pt \halfwidth
0pt \halfwidth
0pt \halfwidth
0pt \halfwidth
0pt \halfwidth
0pt \halfwidth
0pt \halfwidth
0pt \halfwidth
0pt \hsize
Na{\"\i}vely, the Fourier transform $F_N$ requires $O(N^2)$ 
operations.  The classical FFT improves this to $O(N\log N)$.  And 
perhaps the most fundamental result in quantum computation is that 
$F_N$ can be implemented with $O(\log^2 N)$ elementary quantum 
operations [\Shor,\Coopersmith].  Thus we can implement the advection 
step with an $\alpha$-controlled-shift operation 
$|x,\alpha\rangle \mapsto |x+\alpha,\alpha\rangle$, where the shift
(in each direction) is decomposed as in (1).  Slightly less 
efficiently, but with the same $O(\log^2 N)$ complexity, we can 
implement it by acting by a left shift, and then acting by a two step
right shift, conditioned on $\alpha = 1$; this circuit is shown in
Fig.~9.  Thus on a quantum computer we can sample the diffusion 
equilibrium with $O(N\log^2 N)$ elementary operations.  This is an 
improvement by $O(N/\log N)$ over the classical random walk algorithm.

\medskip
\noindent{\bf 6.  Discussion}
\nobreak

\nobreak
\noindent We have seen that a classical physics problem---sampling 
from the equilibrium distribution of a diffusion process---is solvable 
more efficiently with a quantum computer than with a classical one, in
the sense that the QLGA algorithm outperforms the random walk 
algorithm.  The solution utilizes two of the fundamental quantum 
speedups:  First, the quadratic improvement in the number of steps
necessary for a QLGA simulation compared to a random walk simulation
is reminiscent of the quadratic improvement of Grover's quantum search 
algorithm---which he, in fact, describes as diffusion [\Grover].  
Second, the exponential improvement of the quantum Fourier transform
over the FFT [\Shor,\Coopersmith] provides a speedup in the advection
step.  Perhaps the simple problem considered here will inspire 
application of these techniques---or new ones---to speed up 
computation of other classical systems with quantum algorithms.  For
example, a natural generalization would be to analyze diffusion in a
potential, for which a QLGA simulation has already been shown to 
evolve the mean of the distribution quadratically faster than does a 
classical (biased) random walk, in certain cases [\pglga].  Successful 
development of such quantum algorithms should provide additional 
impetus to efforts towards building a quantum computer [\DiVincenzo].

\medskip
\noindent{\bf Acknowledgements}

\noindent I thank Jean-Pierre Boon for his interest in these results 
and for inviting me to present them.  I also thank Bruce Boghosian, 
Peter Coveney, Lov Grover, Manfred Krafczyk, Daniel Lidar, Seth Lloyd, 
Peter Love and Li-Shi Luo for useful conversations, before and during 
the workshop.  This work was supported in part by the National 
Security Agency (NSA) and Advanced Research and Development Activity 
(ARDA) under Army Research Office (ARO) contract numbers 
DAAG55-98-1-0376 and DAAD19-01-1-0520, and by the Air Force Office of 
Scientific Research (AFOSR) under grant number F49620-01-1-0494.

\medskip
\global\setbox1=\hbox{[00]\enspace}
\parindent=\wd1

\noindent{\bf References}
\vskip10pt

\parskip=0pt
\item{[\Feynman]}
\feynman,
``Simulating physics with computers'',
\IJTP\ {\bf 21} (1982) 467--488;\hfb
\feynman,
``Quantum mechanical computers'',
\FP\ {\bf 16} (1986) 507--531.

\item{[\Zalka]}
C. Zalka,
``Efficient simulation of quantum systems by quantum computers'',
\PRSLA\ {\bf 454} (1998) 313--322.

\item{[\Wiesner]}
S. Wiesner,
``Simulations of many-body quantum systems by a quantum computer'',
{\tt quant-ph/9603028}.

\item{[\Lloyd]}
S. Lloyd,
``Universal quantum simulators'',
\Sc\ {\bf 273} (23 August 1996) 1073--1078.

\item{[\BTsim]}
\bogtay,
``Simulating quantum mechanics on a quantum computer'',
\PRD\ {\bf 120} (1998) 30--42.

\item{[\AbramsLloyd]}
D. S. Abrams and S. Lloyd,
``Simulation of many-body Fermi systems on a universal quantum 
  computer'',
\PRL\ {\bf 79} (1997) 2586--2589.

\item{[\FKW]}
M. H. Freedman, A. Kitaev and Z. Wang,
``Simulation of topological field theories by quantum computers'',
{\tt quant-ph/0001071}.

\item{[\OGKL]}
G. Ortiz, J. E. Gubernatis, E. Knill and R. Laflamme,
``Quantum algorithms for fermionic simulations'',
\PRA\ {\bf 64} (2001) 022319/1--14.

\item{[\Shor]}
\shor,
``Algorithms for quantum computation:  discrete logarithms and 
  factoring'',
in S. Goldwasser, ed.,
{\sl Proceedings of the 35th Symposium on Foundations of Computer 
Science}, Santa Fe, NM, 20--22 November 1994
(Los Alamitos, CA:  IEEE Computer Society Press 1994) 124--134;\hfb
\shor,
``Polynomial-time algorithms for prime factorization and discrete 
  logarithms on a quantum computer'',
\SIAMJC\ {\bf 26} (1997) 1484--1509.

\item{[\LidarBiham]}
D. A. Lidar and O. Biham,
``Simulating Ising spin glasses on a quantum computer'',
\PRE\ {\bf 56} (1997) 3661--3681.

\item{[\Yepez]}
J. Yepez,
``A quantum lattice-gas model for computation of fluid dynamics'',
\PRE\ {\bf 63} (2001) 046702.

\item{[\pqa]}
\dajm,
``Physical quantum algorithms'',
UCSD preprint (2001).

\item{[\NielsenChuang]}
M. A. Nielsen and I. L. Chuang,
{\sl Quantum Computation and Quantum Information\/}
(New York:  Cambridge University Press 2000).

\item{[\qcaqlg]}
\dajm,
``From quantum cellular automata to quantum lattice gases'',
\JSP\ {\bf 85} (1996) 551--574.

\item{[\Bennett]}
C. H. Bennett,
``Logical reversibility of computation'',
\IBMJRD\ {\bf 17} (November 1973) 525--532.

\item{[\Toffoli]}
T. Toffoli,
``Bicontinuous extensions of invertible combinatorial functions'',
\MST\ {\bf 14} (1981) 13--23.

\item{[\ADH]}
L. M. Adelman, J. Demarrais and M.-D. A. Huang,
``Quantum computability'',
\SIAMJC\ {\bf 26} (1997) 1524--1540.

\item{[\DeutschJozsa]}
\deutsch\ and R. Jozsa,
``Rapid solution of problems by quantum computation'',
\PRSLA\ {\bf 439} (1992) 553--558.

\item{[\Simon]}
\simon,
``On the power of quantum computation'',
in S. Goldwasser, ed.,
{\sl Proceedings of the 35th Symposium on Foundations of Computer 
Science}, Santa Fe, NM, 20--22 November 1994
(Los Alamitos, CA:  IEEE Computer Society Press 1994) 116--123;
\hfb%
\simon,
``On the power of quantum computation'',
\SIAMJC\ {\bf 26} (1997) 1474--1483.

\item{[\BernsteinVazirani]}
\bv,
``Quantum complexity theory'',
in {\sl Proceedings of the 25th Annual ACM Symposium on the Theory 
        of Computing},
San Diego, CA, 16--18 May 1993
(New York:  ACM 1993) 11--20;\hfb
\bv,
``Quantum complexity theory'',
\SIAMJC\ {\bf 26} (1997) 1411--1473.

\item{[\Grover]}
\grover,
``A fast quantum mechanical algorithm for database search'',
in {\sl Proceedings of the 28th Annual ACM Symposium on the Theory 
        of Computing},
Philadelphia, PA, 22--24 May 1996 
(New York:  ACM 1996) 212--219;\hfb
\grover,
``Quantum mechanics helps in searching for a needle in a haystack'',
\PRL\ {\bf 79} (1997) 325--328.

\item{[\vanDam]}
W. van Dam,
``Quantum algorithms for weighing matrices and quadratic residues'',
{\tt quant-ph/0008059}.

\item{[\qsa]}
M. Hunziker and \dajm,
``Quantum algorithms for highly structured search problems'',
UCSD preprint (2001).

\item{[\BTmulti]}
\bogtay,
``A quantum lattice-gas model for the many-particle Schr\"odinger
  equation in $d$ dimensions'',
\PRE\ {\bf 8} (1997) 705--716.

\item{[\lgbu]}
\dajm,
``Quantum lattice gases and their invariants'',
\IJMPC\ {\bf 8} (1997) 717--735.

\item{[\qlgaI]}
\dajm,
``Quantum mechanics of lattice gas automata:
  One particle plane waves and potentials'',
\PRE\ {\bf 55} (1997) 5261--5269.

\item{[\gauge]}
\dajm,
``From gauge transformations to topology computation in quantum
  lattice gas automata'',
\JPA\ {\bf 34} (2001) 6981--6986.

\item{[\qlgaII]}
\dajm,
``Quantum mechanics of lattice gas automata:  
  Boundary conditions and other inhomogeneities'',
\JPA\ {\bf 31} (1998) 2321--2340.

\item{[\Taylor]}
G. I. Taylor,
``Diffusion by continuous movements'',
\PLMS\ {\bf 20} (1920) 196--212.

\item{[\pglga]}
\dh,
``Parrondo games as lattice gas automata'',
{\tt quant-ph/ 0110028};
to appear in \JSP

\item{[\nogo]}
\dajm,
``On the absence of homogeneous scalar unitary cellular automata'',
\PLA\ {\bf 223} (1996) 337--340.

\item{[\Feynmanlect]}
\feynman, R. B. Leighton and M. L. Sands,
{\sl The Feynman Lectures on Physics}, vol.\ III
(Reading, MA:  Addison-Wesley 1965), Chap.\ 13.

\item{[\FarhiGutmann]}
E. Farhi and S. Gutmann,
``Quantum computation and decision trees'',
\PRA\ {\bf 58} (1998) 915--928.

\item{[\CFG]}
A. M. Childs, E. Farhi and S. Gutmann,
``An example of the difference between quantum and classical random
  walks'',
{\tt quant-ph/0103020}.

\item{[\AAKV]}
D. Aharonov, A. Ambainis, J. Kempe and U. Vazirani,
``Quantum walks on graphs'',
{\tt quant-ph/0012090}.

\item{[\ABNVW]}
A. Ambainis, E. Bach, A. Nayak, A. Vishwanath and J. Watrous,
``One dimensional quantum walks'',
in {\sl Proceedings of the 33rd Annual ACM Symposium on the Theory 
        of Computing},
Hersonissos, Crete, Greece, 6--8 July 2001
(New York:  ACM 2001) 37--49. 

\item{[\Coopersmith]}
D. Coopersmith,
``An approximate Fourier transform useful in quantum factoring'',
IBM Research Report RC 19642 (12 July 1994).

\item{[\DiVincenzo]}
D. P. DiVincenzo,
``The physical implementation of quantum computation'',
{\tt quant-ph/0002077}.

\bye